\documentclass[useAMS,usenatbib]{mn2e}                                                             
\usepackage[totalwidth=515pt,totalheight=660pt,left=1.4cm,right=1.4cm]{geometry}              
\usepackage{graphicx,amssymb,color}                                                         
\usepackage[normalem]{ulem}                                                                    
\input psfig 

\title[H delivery to WDs from comets]
{Hydrogen delivery onto white dwarfs from remnant exo-Oort cloud comets}
\author[Veras, Shannon \& G\"{a}nsicke]{
Dimitri Veras$^{1}$\thanks{E-mail:d.veras@warwick.ac.uk},
Andrew Shannon$^{2}$,
Boris T. G\"{a}nsicke$^{1}$
\\
$^{1}$Department of Physics, University of Warwick, Coventry CV4 7AL, UK
\\
$^{2}$Institute of Astronomy, University of Cambridge, Cambridge CB3 0HA, UK
}

\begin{document}

\date{Accepted 2014 September 26.  Received 2014 September 26; in original form 2014 June 19}

\pagerange{\pageref{firstpage}--\pageref{lastpage}} \pubyear{XXXX} 

\maketitle

\label{firstpage}

\begin{abstract}
The origin of trace hydrogen in white dwarfs (WDs) with He-dominated
  atmospheres is a long-standing problem, one that cannot satisfactorily
  be explained by the historically-favoured hypothesis of accretion
  from the interstellar medium. Here we explore the possibility that
the gradual accretion of exo-Oort cloud comets, which are a rich source of H,
contributes to the apparent increase of trace H with WD cooling age.  
We determine how often remnant
exo-Oort clouds, freshly excited from post-main-sequence stellar mass
loss, dynamically inject comets inside the WD's Roche radius. We
improve upon previous studies by considering a representative range of
single WD masses ($0.52-1.00M_{\odot}$) and incorporating different
cloud architectures, giant branch stellar mass loss, stellar flybys,
Galactic tides and a realistic escape ellipsoid in self-consistent
numerical simulations that integrate beyond 8 Gyr ages of WD cooling.
We find that $\sim 10^{-5}$ of the material in an exo-Oort cloud is
typically amassed onto the WD, and that the H deposits accumulate even
as the cloud dissipates.  This accumulation may account for
the relatively large amount of trace H, $10^{22}-10^{25}$ g, that is determined
frequently among WDs with cooling ages $\ge1$\,Gyr. Our results also 
reaffirm the notion that exo-Oort cloud comets are not the primary
agents of the metal budgets observed in polluted WD
atmospheres.
\end{abstract}


\section{Introduction}

The high surface gravity of white dwarfs (WDs) implies that their
atmospheres are chemically stratified by atomic weight, and the
lightest element floating on top will determine their spectroscopic
appearance \citep{schatzman1958}.  About $80\%$ of the
entire WD population have H-dominated (DA) atmospheres
\citep{koester2013}. The remaining fraction of WDs have lost their
residual H, most likely in a late shell flash \citep{altetal2005}, 
and have He-dominated (DB) atmospheres.

Exquisite compositional analyses from the last few decades have
emphasized that reality often departs from that simple picture:
between a quarter to one half of all WD atmospheres show traces of
metals \citep{zucetal2003,zucetal2010,koeetal2014}, and up to half of
all He-atmosphere WDs contain varying amounts of trace-H
\citep{dufetal2007,vosetal2007,beretal2011}\footnote{The spectra of
  WDs with mixed atmospheres are classified according to the strength
  of the observed lines, e.g. DBA for He-dominated atmospheres with a
  small amount of hydrogen. The presence of metals is indicated by
  adding the letter ``Z'', e.g. DAZ for a metal-polluted H-dominated
  star, or DZ for a featureless cool He-atmosphere with detectable
  metal lines \citep{sioetal1983}.}.  Two important questions then are
(1) from where do the metals originate? (2) what causes the
  apparent changes in the mass of H as a function of age? The
historic answer to both questions is accretion from the interstellar
medium \citep[ISM, e.g.][]{wesemael1979,alcill1980}, but 
this hypothesis has major shortcomings that we discuss below.

Regarding the first question, the diffusion time scale of these
elements in WD atmospheres is short compared to their cooling ages
(\citeauthor{fonmic1979} \citeyear{fonmic1979};
\citeauthor{paqetal1986} \citeyear{paqetal1986};
\citeauthor{koester2009} \citeyear{koester2009}; see also Fig. 1 of
\citeauthor{wyaetal2014} \citeyear{wyaetal2014}).  Therefore, the
spectroscopic detection of elements heavier than H and He does indeed
unambiguously require ongoing or recent accretion. However, the theory
that the accretion originated from the ISM encountered a substantial
number of problems: (1) the small magnitude of the expected accretion
rate from the ISM, (2) the sparse distribution of massive interstellar
clouds, and (3) the existence of metal-rich but H-poor DBZ and DZ WDs
\citep{aanetal1993,frietal2004,jura2006,kilred2007,faretal2010}.
Instead, the presence of metals has been explained by the tidal
breakup and accretion of rocky material
\citep{graetal1990,jura2003,beasok2013,veretal2014a} which subsequently 
form discs that accrete directly onto the WD
\citep{jura2008,rafikov2011a,rafikov2011b,metetal2012,rafgar2012,wyaetal2014}. Further
supporting this thesis are actual detections of dozens of dust discs
\citep{zucbec1987,becetal2005,kiletal2005,reaetal2005,faretal2009,beretal2014},
several gaseous discs \citep{ganetal2006,ganetal2007,ganetal2008,
  gansicke2011,faretal2012,meletal2012} and discs containing both dust
and gas \citep{brietal2009,meletal2010,brietal2012,wiletal2014}.

The second question has received less attention, and is the primary
motivation for this study. Hydrogen, being the lightest element, never
sinks, but it can be subject to convective mixing
\citep{koester1976,vauetal1979}.  Taken at face value, the
  mass of trace H detected in a substantial fraction of He-dominated
  WDs, derived from the analysis of optical spectroscopy and detailed
  envelope models, is found to monotonically increase from
  $M_\mathrm{H}\simeq10^{14}$\,g at a cooling age of a few 100\,Myr to
  $M_\mathrm{H}\simeq10^{25}$\,g at 2\,Gyr\footnote{The Balmer lines
  become increasingly weaker with increasing cooling age / decreasing
  temperature, as most H in the atmosphere is in the ground
  state. Consequently, it is extremely difficult to determine the
  amount of trace H in He-dominated WDs older than $\sim2$\,Gyr.}
(\citeauthor{dufetal2007} \citeyear{dufetal2007};
\citeauthor{vosetal2007} \citeyear{vosetal2007};
\citeauthor{juretal2009} \citeyear{juretal2009};
\citeauthor{beretal2011} \citeyear{beretal2011}; see Fig.~7 of Raddi
et al. 2014). While this distribution of $M_\mathrm{H}$ is
  biased by both physical effects (the depth of the He convection
  zone) and observational limitations, there is strong evidence that
  the oldest WDs carry the largest amount of trace-H
  \citep{dufetal2007}.  \cite{beretal2011} demonstrate that this
age-dependent distribution of trace-H, in particular the
  existence of relatively old WDs with extremely low values of
  $M_\mathrm{H}$, is incompatible with both primordial H left over
  from post-main sequence evolution and interstellar accretion. 
    \cite{stoetal2014} suggest that trace H in old WD atmospheres can
    be used to constrain exo-Oort clouds and the manner in which they
    are dynamically perturbed.

In this paper, we explore the long-term contribution to the H budget
by the accretion of comets from exo-Oort clouds. We outline the
general context of our arguments in Section~2.  In Section~3, we
discuss the interplay of forces relevant to our setup, and why we
utilise numerical simulations in this work. Section~4 provides a
detailed description of those simulations, and Section 5 illustrates
the results.  We conclude with a discussion in Section~6 and a summary
in Section~7.

\section{Comets as external accretors}

\subsection{Lessons from WD pollution}

Much of the recent observational and theoretical work on WD pollution
in evolved planetary systems has focused on the accretion of metals,
for the simple reason that the detection of photospheric metals is a
smoking gun of recent or ongoing accretion, in many cases accompanied
by the simultaneous detection of the circumstellar reservoir of dusty
and gaseous debris. 

The nature of the parent bodies can be inferred from a compositional
analysis of the debris, where the abundance of C is of particular
importance: the mass fraction of C is $\simeq24$\% for comet
Halley, whereas it is $\simeq3.5$\% for CI chondrites, $0.24$\% for
the Sun, and $\simeq0.1$\% for the bulk Earth 
(\citeauthor{jesetal1988} \citeyear{jesetal1988}, 
\citeauthor{alletal2001} \citeyear{alletal2001},
\citeauthor{lodders2003} \citeyear{lodders2003},
\citeauthor{aspetal2009} \citeyear{aspetal2009},
see also Figure 14 of \citeauthor{kleetal2010} \citeyear{kleetal2010}).

Detailed abundance studies have been carried out so far only for a
relatively small number of WDs 
\citep{zucetal2007,kleetal2010,kleetal2011,gaeetal2012,
juretal2012,xuetal2013,xuetal2014}.  However, in all cases, 
the debris was found to be
strongly depleted in C with respect to CI chondrites and hence
incompatible with accretion by material with a composition similar to 
Solar system comets, regardless of the origin of these comets.

An independent method was pursued
by \cite{jurxu2010,jurxu2012} who interpreted the amount of trace-H
in a sample of He-dominated WDs within 80\,pc of the Earth originating
from the accretion of planetary debris, and concluded that the parent
bodies were typically dry, with a water content of at most $\sim
1-10\%$. Exceptions do exist \citep{faretal2013,radetal2014} but
represent distinct outliers \citep[Fig. 7 of][]{radetal2014}, and can
be explained by asteroids which retain water throughout the main
sequence and giant branch phases of the parent star's lifetime
\citep{jurxu2010}.

In conclusion, the observational evidence in the above studies
strongly favours the accretion of rocky, asteroid-like debris over
exo-Oort cloud comets. However, \textit{one bias that all these studies have in
  common is that they only addressed relatively young WDs}, with
cooling ages $\la300$\,Myr. The accretion of these comets may be less
frequent and more challenging in terms of direct detection when
compared to asteroids, but could still have a noticeable and
observable long-term effect on the properties of WDs.

\subsection{Lessons from the Solar system}

Observed collisions of small bodies with the Sun, which is an almost 5
Gyr-old main sequence star, may help us understand the collision
frequency with small bodies at $\sim 5$ Gyr-old WD cooling ages.  What
we see locally, however, does not necessarily agree with our current
understanding of metal-polluted WD systems.

Comets impact the Sun frequently.  In fact, coronographs like those
part of SoHO's LASCO reveal that a comet grazes the Sun every few
days, with a total of about 2400 grazers from 1996 to 2008
\citep{lametal2013}.  However, about $90\%$ of these comets are from a
single family (Kreutz) and have similar pericentres (about two Solar
radii) and other orbital parameters.  Further, the number of comets
counted is a strong function of their brightness \citep{sekkra2013},
and the observations do not provide physical parameters such as mass
and radius.  Nevertheless, large pieces of the Kreutz comet have been
observed from the ground for centuries, and the estimated mass of the
progenitor is at least $0.5 \times 10^{20}$ g \citep[Table 1
  of][]{knietal2010}.  By comparing this value to the heavy element
mass in DBZ WDs that was accreted over the last 1 Myr or so years
\citep[Fig. 9 of][]{giretal2012}, we see that the Kreutz comet by
itself could explain the lowest accreted masses within just the last
few hundred years.  This result is independent of the comet's
fragmentation frequency, number and type (sublimation-, ablation- or
explosion-dominated; see \citealt*{broetal2011}).

Contrastingly, asteroids infrequently graze the Sun.  The grazing
or collisional timescale is on the order of Gyr \citep{minmal2010};
even asteroids in unstable resonances have collisional timescales
of $\sim$ Myr (Table 1 of \citealt*{glaetal1997}).

\subsection{Our new contribution}

This disagreement between observations inside and outside of our Solar System 
acts as a secondary motivation for our study, in addition to explaining the 
increasing mass of trace-H found among He-dominated WDs with time. 

In a pioneering early investigation, \cite{alcetal1986} estimated that a 
exo-Oort cloud comet hits a WD about every $10^{4}$ years, 
a value subsequently adopted by 
studies such as \cite{debsig2002} and \cite{zucetal2007}.  
In an extension of \cite{alcetal1986}, \cite{paralc1998} used the same 
underlying mass loss rate, but with an additive fictitious force to model 
asymmetric mass loss.  They found that for an asymmetry which imparts an 
impulsive kick to the star on the scale of hundreds of m/s, the comet 
accretion rate is reduced. \cite{stoetal2014} also
adopts impulsive kicks, yielding accretion rates which span several
orders of magnitude depending on the adopted kick velocities and 
pericentres (see their Fig. 1).

In this work, we extend the study of \cite{alcetal1986} by using a
numerical integrator which includes flybys, tides, a realistic escape
ellipsoid, stellar evolution profiles from a different dedicated
stellar evolution code, a relevant range of progenitor stellar masses,
and different radial profiles of exo-Oort cloud comets.  Given the
uncertainty in the mass of exo-Oort clouds and the limited resolution
of numerical integrations, our primary output will be a fraction of
comets which reach the Roche radius of the WD. From this value, one
can determine a mass accretion rate by making assumptions about the
mass of the comets and cloud.

\section{Relevant forces}

Throughout a star's lifetime, comets in an exo-Oort cloud are perturbed by both Galactic tides and stellar flybys.  If a comet exits the star's gravitational sphere of influence, then the comet has escaped into the interstellar medium.  Similarly, comets already in the interstellar medium may enter the star's gravitational sphere of influence.  The sphere of influence is technically not a sphere, but rather a triaxial ellipsoid whose three axes are given analytically by equations (21-22) of \cite{veretal2014b}.

We do not know the extent of the reservoir of interstellar comets.  However, a few dedicated studies provide some estimates.  Table 2 of \cite{moretal2009} indicates that the number density of comets outside of an exo-Oort cloud is about $10^7 - 10^9$ comets per cubic parsec.  If the cloud itself contains about $10^{11}$ comets in an annulus extending from $10^4$ au to $10^5$ au, then the density within the cloud is $5 \times 10^{10}$ comets per cubic parsec.  Hence, the cloud itself is a few orders of magnitude more dense than its surroundings.  Also, \cite{jura2011} importantly notes that no observed Sun-grazing comets have hyperbolic orbits, meaning that they all originated from inside of the Solar System.  He places a specific constraint on interstellar comets with radii between 10 m and 2 km, claiming that they compose less than 1\% of all interstellar oxygen.  Further, interstellar comets have relative velocities on the order of 10 km/s, whereas Oort cloud comets instead have velocities which are at least one order of magnitude less. Hence, the star's gravity sets the path of its own comets, and bends interstellar comets towards it significantly only if they come within a few au. Overall, all these arguments help us justify neglecting capture of interstellar comets in this work.

When the star leaves the main sequence and evolves along giant branches, another important force affects the orbit of these comets: stellar mass loss.  When the star becomes a WD, mass loss effectively ceases.  The timescale for cometary orbits to change due to mass loss is much shorter than the timescale for changes due to the Galactic tide \citep{veretal2014b}.  Consequently, these forces are effectively decoupled.  

Given their large orbits ($10^4 - 10^5$ au in semimajor axis), all long-period comets will nonadiabatically respond to stellar mass loss \citep{veretal2011}.  By {\it nonadiabatic} we mean that the behaviour is not simply described by a constant-eccentricity orbit expansion, as would be true for objects (such as planets) within a few hundred au of their parent stars.  A comet's actual response to the mass loss is well described by a set of orbital element evolution equations in both the isotropic mass loss case (equations 35-37 of \citealt*{omarov1962} and equation 29 of \citealt*{hadjidemetriou1963}) and anisotropic mass loss case \citep[equations 18-23 and 34-38 of][]{veretal2013b}.  Both cases demonstrate that cometary orbits will evolve in a non-obvious manner, featuring both decreases and increases in their eccentricities depending on their other orbital parameters.

The importance of anisotropy in mass loss was highlighted by \cite{paralc1998} but not fully described with orbital elements in terms of the variable mass flux at different positions on the stellar surface until \cite{veretal2013b}.  The latter study highlights two important points: (1) if anisotropy exists, its importance scales as the square root of the semimajor axis, and (2) physically-sensible scaling laws for mass ejecta as functions of stellar latitude and co-longitude predict no net anisotropic effect whatsoever (their equations 51-53).  These same laws, however, fail to reproduce observations which feature clearly asymmetrical structures like planetary nebulae \citep[e.g.][]{kimetal2008}.  Hence, our knowledge of stellar physics still remains insufficient to adopt any particularly representative asymmetric mass loss prescription along the giant branches.

The general equations of motion for the two-body problem with mass loss and Galactic tides have no known analytic solution.  Hence the correlation, for example, between initial conditions and escape from the Oort cloud is complex \citep[see Figs. 10-11 of][]{verwya2012}.  However, useful dependencies can be extracted by analysing their form with orbital elements.  For example, \cite{veretal2011} prove that the pericentre of a comet can never decrease while the parent star is losing mass isotropically.  \cite{heitre1986}, \cite{brasser2001} and \cite{vereva2013} show that in the Solar neighbourhood, a comet whose orbit is coplanar with the Galactic disc robustly resists changes to its semimajor axis and eccentricity because the Galactic vertical tide vanishes.

Nevertheless, these relationships may be significantly modulated in the presence of a flyby star.  \cite{kairay2014} particularly emphasise the importance of including stellar flybys for calculations of extremely close pericentre passages.  The geometric details and impact parameters of flybys around stars other than the Sun is unknown.  Even for the Sun, flyby trajectories can be computed only for a few Myr and only for a fraction of the stars in the Solar neighbourhood \citep{jimetal2011}.  Such uncertainty is compounded by the prospect of flyby binaries or flyby planetary systems \citep{vermoe2012}.  A flyby can affect an orbiting planet in a variety of ways.  Hence, in concert with Galactic tides, perturbations can force back down the pericentre of a comet which is raised by stellar mass loss.

\section{Numerical simulations}

This investigation determines the extent of the effects detailed in the last section, with important implications for WD pollution.  The above discussion, coupled with the fact that energy is not conserved in mass-losing systems, suggests that a purely analytic approach is not feasible. Therefore, we turn to numerical simulations.

\subsection{Integration procedure}

In order to self-consistently model mass loss, Galactic tidal perturbations and multiple 
stellar flybys in an $N$-body integrator, we have heavily modified the integrator suite
{\tt Mercury} \citep{chambers1999}.  The modifications include the following.

\begin{itemize}
\item{We incorporate stellar mass loss into the code by splicing in-between Bulirsch-Stoer 
timesteps, which although is perhaps not necessary for test particle systems like ours here
\citep{veretal2011}, significantly increases the accuracy for multiple massive objects
\citep[see Fig. 1 of][]{veretal2013a}.
}

\item{When the star is not losing mass, the standard non-conservative Bulirsch-Stoer 
integrator is still used because the perturbation on a comet due to a flyby may be arbitrarily large.
When perturbations are large, symplectic integrators may become inaccurate.
}

\item{Stellar flybys are modelled as perturbative accelerations to all of the comets and the
parent star.  A new flyby is introduced when a probability threshold is reached after an 
individual timestep.}

\item{We incorporate into the code the same prescription for Galactic tides as 
in \cite{vereva2013} and \cite{shaetal2014a,shaetal2014b} and assume
our modeled systems reside in the Solar neighbourhood, specifically at 8 kpc from the
Galactic centre.  We include both horizontal and vertical tides, and contributions from
an exponential disc, a Hernquist bulge, and a cored isothermal halo.}

\item{Because pollution likely arises from disrupted bodies forming discs or rings around
the star, we replace the actual WD radius with the WD Roche radius.  This change may significantly
affect results relating to close encounters with the star \citep{musetal2014}.}

\item{We explicitly incorporate the Hill escape ellipsoid \citep{veretal2014b} into {\tt Mercury} 
for every system sampled to flag ejection, thereby allowing us to quantify escape in preferential
directions. This ellipsoid is dependent on both the stellar mass and the Galactic tidal prescription used.}
\end{itemize}

We use a slightly simplified version of the local field star mass function presented
in \cite{paretal2011} (see their Fig. 7), with a Solar neighbourhood-specific
spatial stellar density of $n_{\star} = 0.392$ pc$^{-3}$ and a stellar flyby distribution which 
is uniformly distributed in log mass such that $-1.2 \le \ln\left(M_{\rm b}/M_{\odot}\right) \le 0.1$. Here, $M_{\rm b}$ refers to the mass of the stellar flyby.
We also simplify our model by adopting an encounter velocity of $V_{\rm b} = 45.7$ km/s,
which corresponds with the expected mean velocity in the Solar neighbourhood
\citep{garetal2001}.

Incoming stars are initially placed at a distance $r_{0}$ in a randomly
chosen direction from the Sun.  They perturb all the bodies of the
simulation as point masses.  Flyby stars are assumed to move in straight
lines (i.e. with hyperbolic eccentricities of infinity) with speed
$V_{\rm b}$, also in a randomly-chosen direction (with $\vec{V}_{\rm b}\cdot \vec{r}_{0} < 0$
so that the star initially travels towards the Sun)\footnote{Our
approximation is robust because for curved trajectories, the pericentre
velocity of the flyby stars closely mimics $V_{\rm b}$ unless the flyby
star has an unusually slow and close orbit.  For a stellar flyby eccentricity
$e_{\rm b}$, the pericentre velocity strays
by more than 10\% only in the special case of $1 < e_{\rm b} \lesssim 10.5$
and strays by over a factor of two only if $1 < e_{\rm b} < 5/3$.}.
Stars spend an average time $r_{0}/V_{\rm b}$ within $r_0$ of the Sun.
Consequently, during every integration timestep, of length $\Delta t$,
we introduce a new star with probability

\begin{equation}
P_{\rm b} = 
\frac{4\pi}{3}
n_{\star}
r_{0}^2
V_{\rm b}
\Delta t
\end{equation}

\noindent{}so that the local stellar density is $n_{\star}$ on long timescales.
The coefficient of $\frac{4}{3}$ is from our use of spherical geometry, rather than the
cylindrical geometry of, e.g., \cite{steshu1988}.  The value of $\Delta t$ is determined
by a Bulirsch-Stoer accuracy parameter of $10^{-13}$.  We chose $r_0 = 2.5 \times 10^5$ au,
as encounters at larger distances are not expected to significantly affect the
evolution.

\begin{figure*}
\centerline{
\psfig{figure=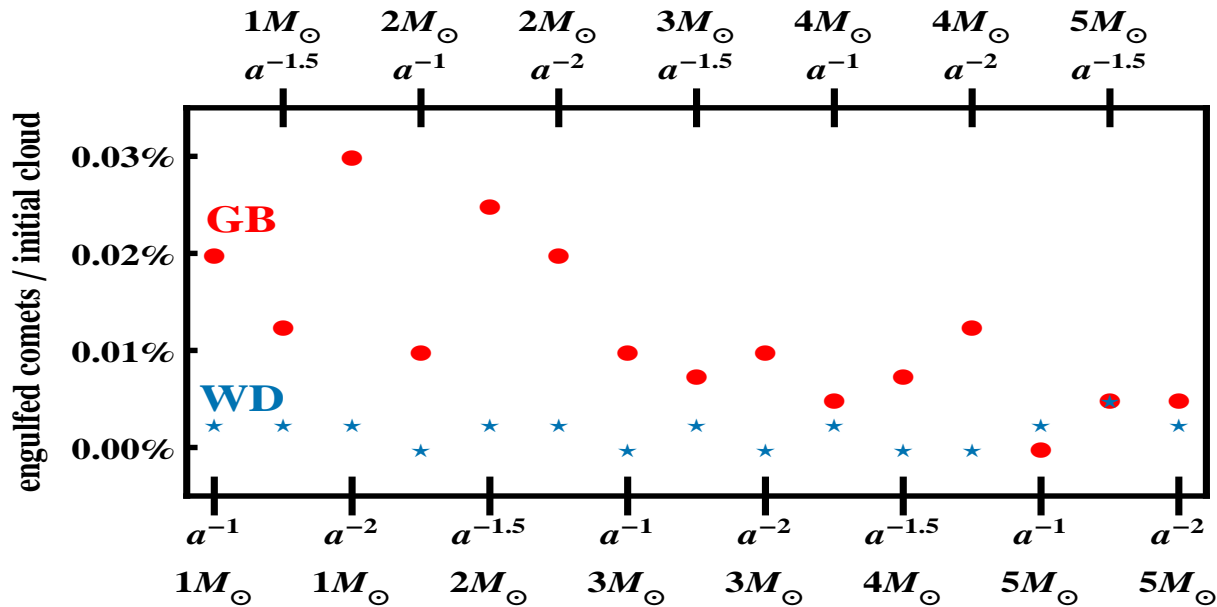,height=9.5cm,width=17cm}
}
\caption{The fraction of comets in exo-Oort clouds which enter the Roche radius of the star when it is a giant branch star (GB; red filled dots) or a white dwarf (WD; blue stars).  The results of all 120 simulations of 5000 comets each are presented, binned according to the main sequence stellar mass and semimajor axis profile of comets (shown on the top and bottom axes).  About 87\% of all engulfments by the star occur between the end of the main sequence and the beginning of the WD stage (the mass-losing phase). Hence, comets are expect to hit the WD at a rate of about one per $10^4$ yr. 
}
\label{destroyed}
\end{figure*}

\subsection{Initial conditions}

We perform 5 sets of 24 simulations of exo-Oort clouds lasting for 10 Gyr starting
from the end of the main sequence.  Each set contains a different stellar progenitor
main sequence mass ($1 M_{\odot}$, $2 M_{\odot}$, $3 M_{\odot}$, $4 M_{\odot}$, $5 M_{\odot}$).  The stars
are evolved in mass and radius according to the stellar evolution code {\tt SSE}
\citep{huretal2000}, which assumes isotropic mass loss (see Section 2). The stars are assumed to have Solar metallicity. The duration
of the mass loss phases for the stellar progenitors are, respectively, 1.431 Gyr, 331.6 Myr, 99.93 Myr,
36.11 Myr, and 17.51 Myr, and their eventual WD masses are 0.52$M_{\odot}$,
0.64$M_{\odot}$, 0.75$M_{\odot}$, 0.87$M_{\odot}$, and 1.00$M_{\odot}$. For more details on the evolution of these particular stars see \cite{veretal2013a}.

We model exo-Oort clouds with 5000 comets each, where each comet is treated as a test particle.  These 
comets have semimajor axes which are distributed
between $10^4$ au and $10^5$ au according to three power law distributions, $a^{-1}$, $a^{-3/2}$ and $a^{-2}$.
Each of the distributions accounts for 8 of the 24 simulations in each set.
The comets' eccentricities, mean anomalies, longitudes of ascending node and arguments of pericentre
are drawn from a uniform random distribution.  The eccentricities of Oort cloud comets were
originally assumed to be isotropically distributed \citep{oort1950} because even if these comets
once harboured high eccentricities, these eccentricities could have subseqently been damped through
stellar flybys and Galactic tides.  Because we assume the star is spherical, the comet's
inclination is measured with respect to the Galactic plane.  This value
crucially dictates the strength of the Galactic tide.  Therefore, we have sampled comets
from a $\sin{i}$ distribution for $i = 0^{\circ}-90^{\circ}$ and $i = 180^{\circ}-270^{\circ}$,
and from a $\sin{\left(90^{\circ}-i\right)}$ distribution for $i = 90^{\circ}-180^{\circ}$ and 
$i = 270^{\circ}-360^{\circ}$.  Doing so helps prevent us from undersampling the phase space
around the inclination values of $90^{\circ}$ and $270^{\circ}$, which would cause the greatest 
orbital variation due to tides.

\section{Results}

\subsection{Exo-Oort cloud comets as a source of H-accretion}

Our key result is illustrated in Fig. \ref{destroyed}.  This figure
bins all of the simulations in 15 groups, for different radial comet
profiles and progenitor main sequence stellar masses.  For each bin,
the figure displays the fraction of comets which have entered the Roche 
radius of the star throughout the simulations.  The characteristic value is $\sim
0.01\%$.  However, about $87\%$ of these comet engulfments occur when
the star is on the giant branch.  Therefore, the characteristic
engulfment fraction for WDs is $\sim 0.001\%$.  These values may be
higher by a factor of a couple for main sequence masses of $1-2
M_{\odot}$.

Consequently, no more than $10^{-5}$ of an initial exo-Oort cloud
should pollute the average WD.  Therefore, for a typical 
main sequence exo-Oort
cloud composed of $10^{11}$ comets, at most $10^6$ comets could ever
reach the WD over $10^{10}$ years.  Hence, one comet intersects with
the WD Roche radius every $10^4$ years, which is the same order of
magnitude predicted by \cite{alcetal1986}.  Figure \ref{destroyed}
indicates that this result is relatively insensitive to stellar mass
or the semimajor axis distribution of comets.

We can place these values in context by considering Fig. 7 of
\cite{radetal2014}. Every WD on this plot which has
accumulated more than $10^{22}$ g of H
can be explained in our model as accretion from an exo-Oort cloud
that is at least as massive as $10^{28}$ g. Many of the older WDs
contain much more H, up to about $10^{25}$ g.  These stars have
accumulated their H from either rocky material like asteroids, particularly 
massive exo-Oort clouds, or a few particularly massive comets from
an exo-Oort cloud (the mass distribution of the Solar system's 
long-period comets is top-heavy; \citealt*{fersos2012}).

Exo-Oort cloud masses are largely unconstrained observationally, 
although reasonable estimates can be obtained to within a couple 
orders of magnitude (as in e.g., \citealt{moretal2009}).  
We assume that these clouds have masses that are distributed about 
some median value which is comparable to that in the Oort cloud.  
Further, the assumption that 
the Oort cloud is typical is congruent with observed planetary and 
protoplanetary systems and models of Oort cloud formation. Regardless, 
WDs are likely to accrete a large number of comets throughout their 
very long existence, and this process could significantly contribute 
to observed trace H in old He-atmosphere WDs.  In addition, 
water-rich asteroids (e.g. \citealt{faretal2013,radetal2014}) 
provide an extra source of H; accretion from the interstellar 
medium, exo-Kuiper belts, and `mini-exo-Oort clouds' \citep{rayarm2013} 
may contribute to some extent as well.  However, 
the fraction of water-ice which can survive post-main-sequence
evolution may be a strong function of both size and distance
\citep{steetal1990,jurxu2010}.

\begin{figure}
\centerline{
\psfig{figure=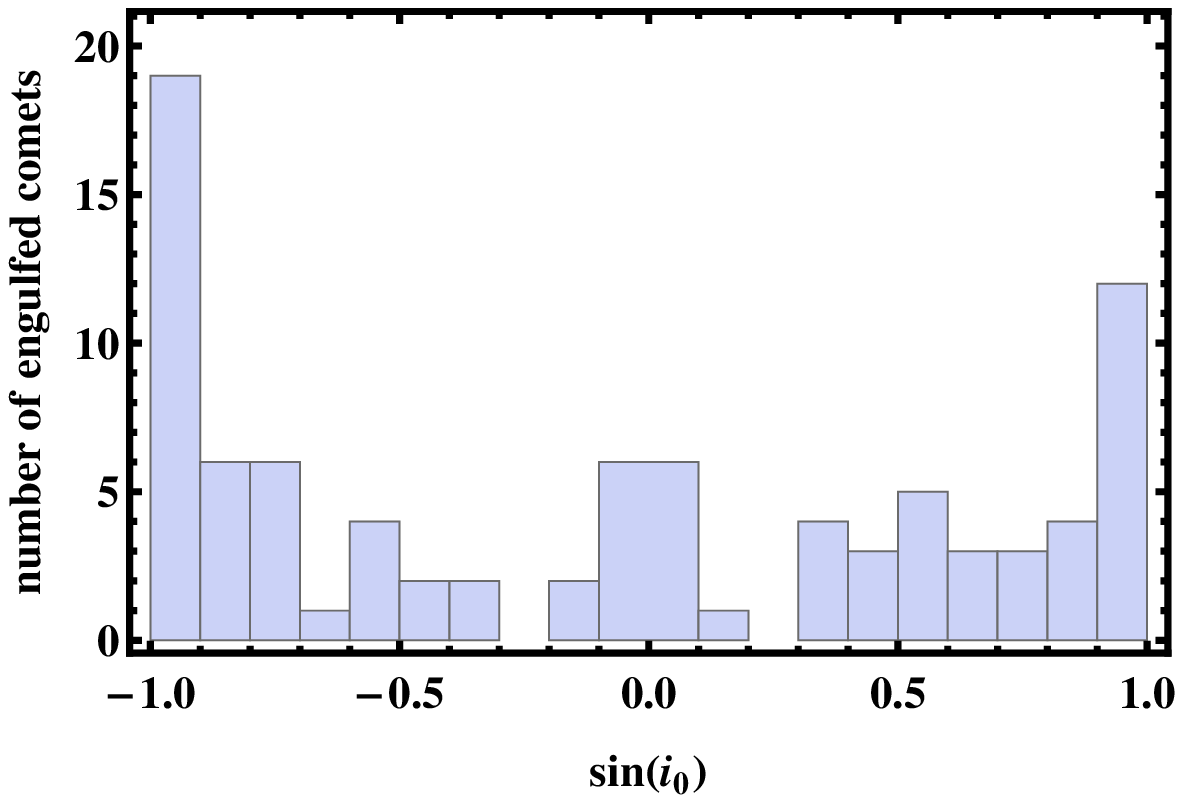,height=5.5cm,width=8.5cm} 
}
\centerline{
\psfig{figure=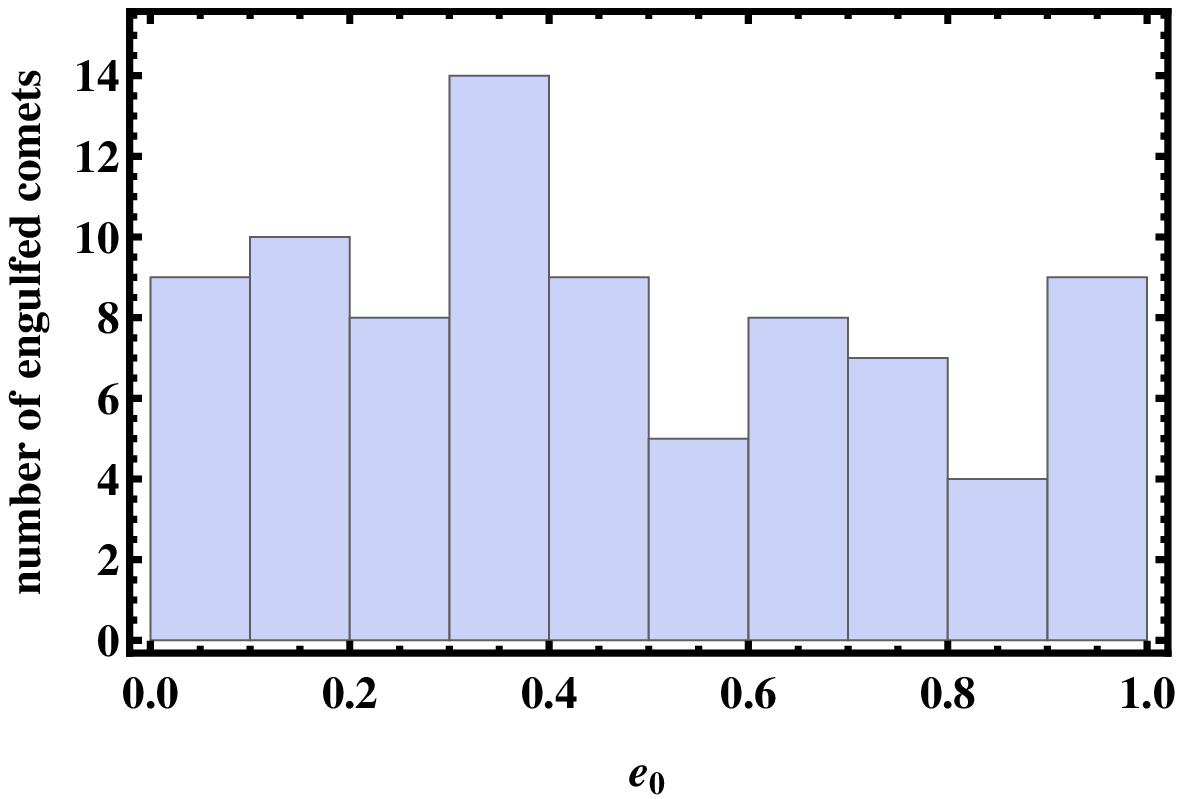,height=5.5cm,width=8.5cm} 
}
\centerline{
\psfig{figure=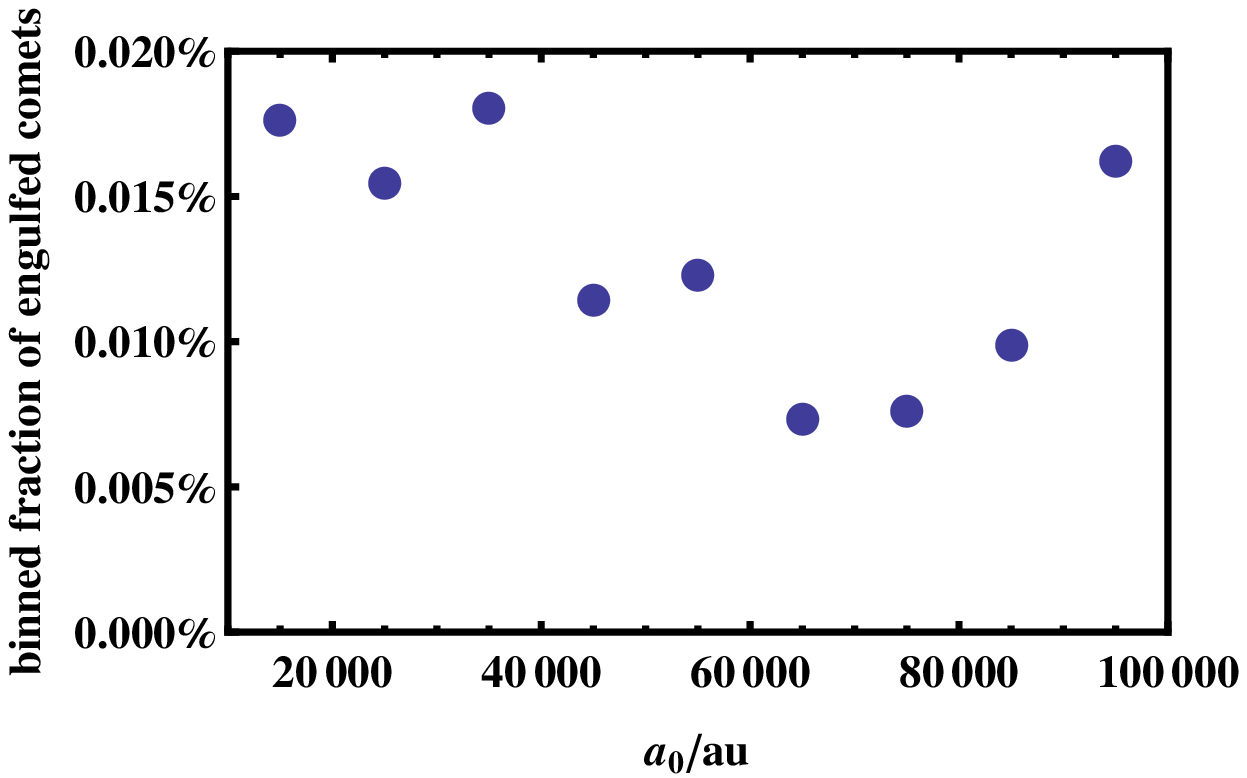,height=5.5cm,width=8.5cm} 
}
\caption{
The initial distributions of the sines of the inclination, 
the eccentricities, and semimajor axes of all exo-Oort cloud 
comets which were engulfed across all simulations.  The 
initial inclination distribution of the engulfed comets largely 
follows the initial inclination distribution of all comets,
and the relatively flat eccentricity distribution exhibits a
similar correlation. The nonzero values in each semimajor axis
bin demonstrate that engulfment occurred across the range sampled.
}
\label{ei}
\end{figure}

\subsection{Exo-Oort cloud comets as a source of metal pollution}


As some Oort cloud comets do contain metals in addition to H, we now consider the potential
contribution of exo-Oort cloud comet accretion to the metal budget of WD atmospheres.
We preface the computations with the caveat that the fraction of metals versus
H in comets is largely unknown.

When a WD accretes metals, the metals will remain in the
convective layer for as little as a few days, to as long as a few
million years, depending on the composition of the envelope, and the
temperature of the WD. As the depth of the convection zone
can be calculated, a measurement of the abundances of metals in the
atmosphere can be understood in terms of the total mass of metals in
the convection zone, and thus, a lower bound for the total amount of
metals accreted. Model inferences from observations find that 
the total mass of metals
within deep convection zones of DBZ WDs is $10^{20-25}$ g
\citep{giretal2012}~for those WDs with detectable amounts of
pollution. Combined with the time for metals to sink out of the
convective zone (see e.g. \citealt*{paqetal1986} and \citealt*{koester2009}), 
a mass accretion rate of exo-Oort cloud comets can be derived.

\cite{wyaetal2014} performed this calculation.  In doing so, they also
observationally debiased the values from \citet{giretal2012}, allowing
them to fit a population model of small bodies accreted by WDs 
to the measurements of the total mass of metals within the deep
convective layers. They were able to fit the observations with
a log-normal distribution with a mean rate of $\dot{M}
\approx 10^{8}~\rm{g}~\rm{s}^{-1}$, and a standard deviation of
$\sigma \approx 1.3$ dex.  They also fit the size-number distribution
of the small bodies as

\begin{equation}
n\left(>R\right) = n_0R^{1-q}
\end{equation}

\noindent{}for the number of small bodies $n$~with radius greater than
$R$, so that the population satisfies $q \approx 3$. This value is
very close to the value found for long period comets \citep[$q = 3.2
  \pm 0.2$,][]{fersos2012}, the asteroid belt and scattered disk
\citep[$q \approx 2.5$,][]{botetal2005,viletal2012}, trans-Neptunian
objects as a whole \citep[$q \approx 4$,][]{truetal2001,fraetal2008},
and inferred for extrasolar small body populations \citep[$q \sim
  4$,][]{shawu2011}.  Hence, a population of small bodies such as
comets is compatible with the polluter population. If we assume that 
distributions of extrasolar comets and asteroids have the same $q$
values as those in the Solar system, then $q\simeq3$ does not allow
us to differentiate between any of these populations of planetary bodies.

However, producing the mass accretion rate found in \citet{wyaetal2014} by 
long-period comets is far more difficult. Our calculations found that
roughly $10^{-5}$~of the primordial population was accreted by the WD
over roughly $10^{10}~\rm{yrs}$. Consequently, we obtain a mass
accretion rate of $10^{-15}~\rm{yr}^{-1} = 3 \times
10^{-23}~\rm{s}^{-1}$~of the primordial population. This rate would
demand a mean Oort cloud mass of $\sim 3 \times 10^{30}~\rm{g}
\approx 500 M_{\oplus}$, which is $10-100$~times the estimated mass of
the Oort cloud
\citep{weissman1996,donetal2004,francis2005,brasser2008}.  Although
the efficiency with which objects are implanted into the Oort cloud
may be higher for systems with lower-mass planets \citep{lewetal2013}
and can depend on the stellar environment in which it is formed
\citep{kaiqui2008,braetal2012}, even a formation channel with $\sim
100\%$~efficiency would still require a mass budget of $\sim 10$~times
the Minimum Mass Solar Nebula
\citep{weidenschilling1977,hayashi1981}. Five hundred earth masses of
solids also exceeds by a factor of $\sim 10$~the total mass budget of
solids available to a typical protoplanetary disk \citep{wilcie2011}.

Although WD progenitors are somewhat more massive than the Sun on
average, the mass of protoplanetary disks increases only linearly with
the stellar mass \citep{wilcie2011}. Thus while a $500 M_{\oplus}$~Oort cloud
might be possible under the most optimistic of assumptions, it seems
very difficult to produce such clouds as the mean of the
population. When one considers the clouds which are one standard
deviation above the mean value of ($\sim 10^4 M_{\oplus}$)~even
optimistic investigators would likely conclude that long-period comets
cannot be the primary agents responsible for the observed metal
pollution of WDs~--~which is consistent with the conclusions
drawn from the abundance studies of the circumstellar debris (Sect. 2.1).

However, the accretion rate of $3 \times 10^{-23}~\rm{s}^{-1}$~of the
primordial population, when applied to estimated mass of the Oort
cloud ($\sim 10 M_{\oplus}$)~yields a mass accretion rate of $\sim
10^6~\rm{g}~\rm{s}^{-1}$, just shy of the lowest detected rates in
\citet{wyaetal2014}. Thus, if the Solar System's Oort cloud is
typical, we should expect to find WDs polluted by
cometary material in the future from the high end tail of the exo-Oort
cloud mass distribution tail.

\subsubsection{Link to initial orbital properties}

Now we take a closer look at the how the engulfment of 
exo-Oort cloud comets is
correlated to their initial orbital properties in Figure \ref{ei}.
The figure shows that the distribution of the initial inclinations and
eccentricities of the engulfed comets largely mirrors the
initial distributions of all comets.  Although comets with smaller semimajor
axes are preferentially engulfed by the WD, the nonzero bins up to 
$a_0 = 10^5$ au demonstrate that the entire range
of sampled initial semimajor axes yield engulfed comets.

Some combination of Galactic tides and flybys could play a significant
role in particular systems. The systems in the two bins closest to 
$\sin{i_0} = \pm 1$ would be most affected by vertical tides.  The 
nonzero bins surrounding $\sin{i_0} = 0$ raise the possibility
that the radial Galactic tides by themselves are strong enough to produce
engulfment, or that flybys are the dominant driver of engulfment
in those cases.
\cite{kairay2014} note that achieving a pericentre which causes
star-star collisions from wide binaries is primarily due to external
stellar flybys, because tidally changing the pericentre becomes more
difficult as the pericentre approaches zero.  Their pericentre value
of about a Solar radius is similar to the Roche radius of a typical WD
that we adopt here, and so their result should hold here also.
However, their statement is true only for vertical tides, as the
dependence of eccentricity on radial tides is complex
\citep[e.g. equation A2 of][]{vereva2013}.  In fact, Fig. 8 of
\cite{levetal2006} shows how radial tides can drive the pericentre
down by a factor of 2.  

\subsubsection{Dissipation of the exo-Oort cloud}

During and after the giant branch phases, the exo-Oort cloud
is decimated.  Although our WD Roche
radius collision rate is similar to the overall collision rate of
\cite{alcetal1986}, we differ on our estimates for the retention of
Oort cloud comets during mass loss.  Our results for exo-Oort cloud
retention are significantly more pessimistic.  \cite{alcetal1986}
suggest that over half of the long-period, highly eccentric comets are
retained during mass loss.  We find that over half of our exo-Oort
clouds are retained during mass loss for only a subset of our
simulations with a progenitor main sequence stellar mass of $1
M_{\odot}$.  Even for those simulations, less than 15\% of the Oort
cloud is retained after 2 Gyr of WD cooling.

\begin{figure}                                               
\centerline{                                                                                            
\psfig{figure=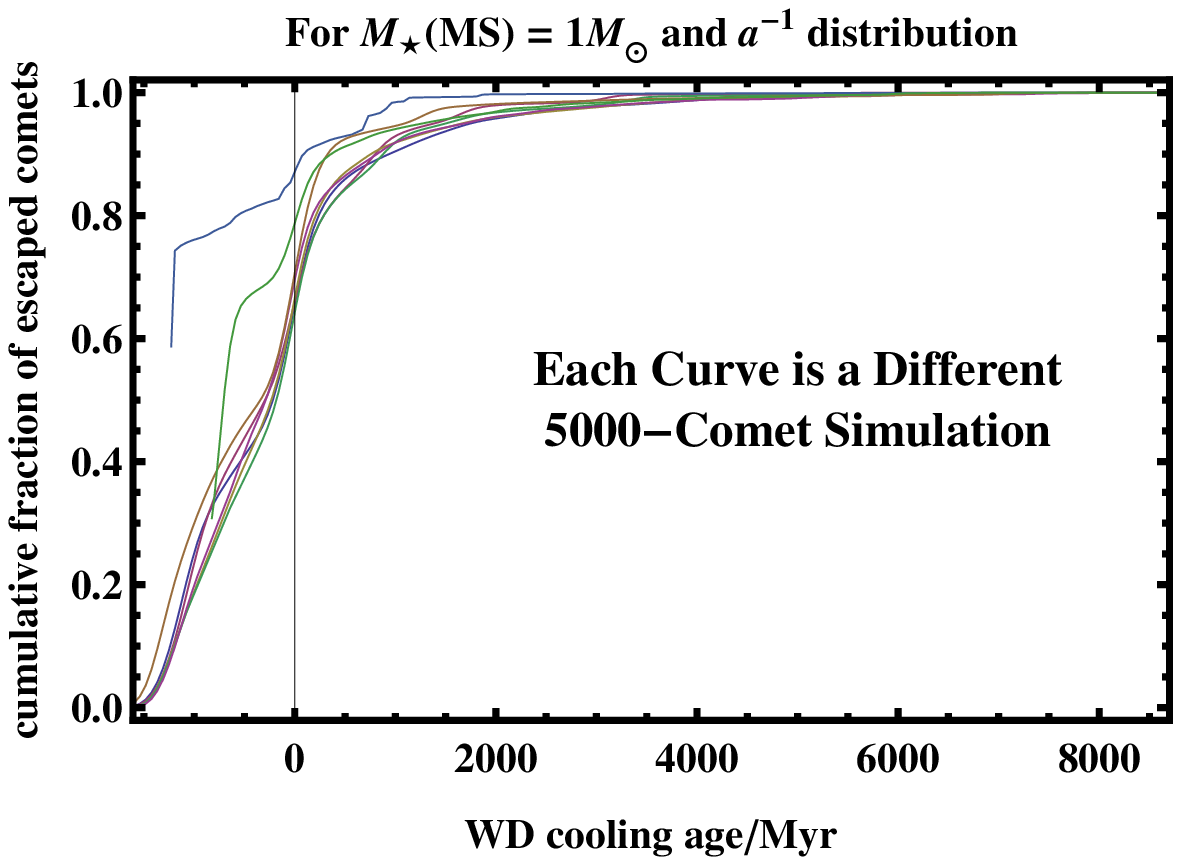,height=5.4cm,width=8.5cm}                                                    
}                                                                                                       
\centerline{ }                                                                                          
\centerline{                                                                                            
\psfig{figure=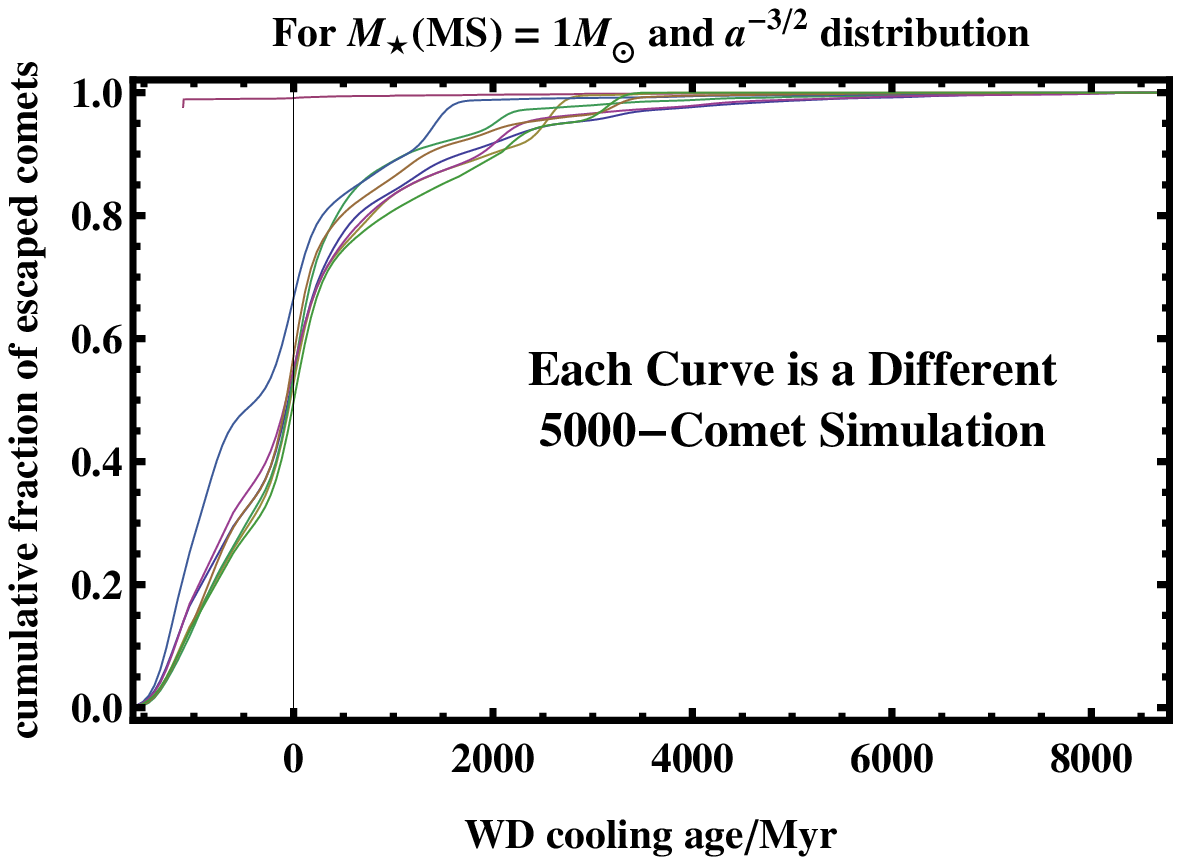,height=5.4cm,width=8.5cm}                                                    
}                                                                                                       
\centerline{ }                                                                                          
\centerline{                                                                                            
\psfig{figure=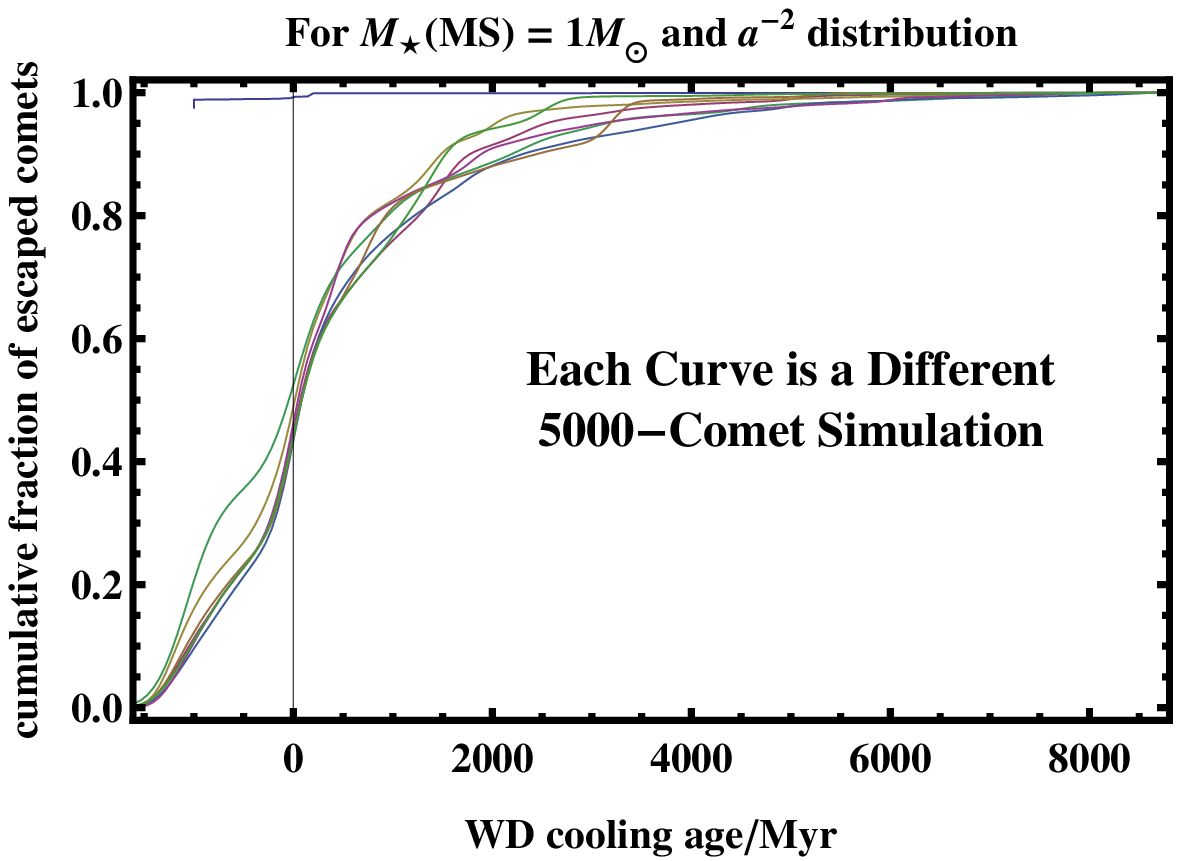,height=5.4cm,width=8.5cm}                                                    
}                                                                                                       
\caption{                                                                                               
Cloud dissipation: the cumulative distributions of escaped exo-Oort cloud                               
comets as a function of time                                                                            
for all simulations with a $1M_{\odot}$ main sequence progenitor                                        
stellar mass.  Each plot corresponds to a different initial semimajor                                   
axis distribution of comets and contains 8 curves.  All curves illustrate                               
that the exo-Oort cloud is at least 95\% depleted by WD cooling ages of 5 Gyr.                          
For earlier cooling ages, typically less than half of the comets remain in                              
the cloud.  For higher progenitor stellar masses, the cloud depletes at                                 
a faster rate.                                                                                          
}                                                                                                       
\label{escape}                                                                                          
\end{figure}                                             

The primary reasons for the disagreement are model-dependent.  Some
factors are our use of a Hill escape ellipsoid rather than an escape
sphere, the different orbital distribution of comets at the start of
mass loss, the presence of flybys during mass loss, and potentially
the different stellar evolution profiles we use for a variety of
stellar masses (although \citealt*{alcetal1986} do consider a wide
range of total mass lost and adiabaticity).  The key factor perhaps is
that highly nonadiabatic mass loss at the tip of the asymptotic giant
branch typically raises the semimajor axis of a comet by a factor
which is higher than a few \citep{veretal2011}, and consequently
ejects the comet either during mass loss or soon after (within one
orbit).  The combination of orbital expansion and Hill ellipsoid
shrinkage when a star becomes a WD means that 51\%-88\% of the volume
of the entire main sequence Hill ellipsoid is a danger zone that will
later eject any occupants \citep[see Section 5.1 and Fig. 6
  of][]{veretal2014b}.

We illustrate the time-dependent ejection of the exo-Oort clouds for
the lowest stellar mass case in Fig. \ref{escape}.  In all other
cases, over 80\% of the cloud is ejected when the star begins life as
a WD.  The figure displays three sets of 8 cumulative overlaid
histograms of the ejection fraction of the original exo-Oort cloud
with respect to the time when the star becomes a WD.  The survival
time of the cloud correlates with the steepness of the initial
semimajor axis profile.  When comets are initially concentrated away
from the edge of the Hill ellipsoid, they are likely to survive for
longer.  The oldest metal polluted WDs, with cooling ages of about 5
Gyr \citep{faretal2011,koeetal2011}, can
expect to have retained no more than 5\% of the cloud which existed at
the beginning of the giant branches.

\begin{figure}
\centerline{
\psfig{figure=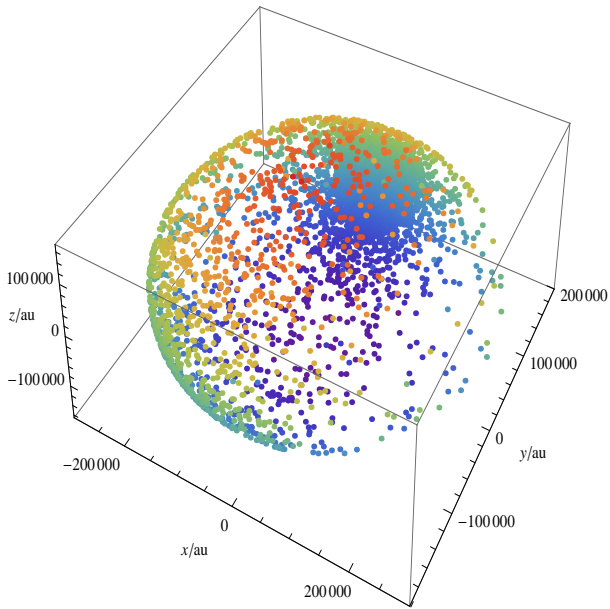,height=8cm,width=8cm}
}
\caption{Patchy escape due to an intrusive stellar flyby: the Cartesian $(x,y,z)$ locations where comets escape the Hill ellipsoid for a single 
5000-comet 10 Gyr simulation with a stellar progenitor mass of $1 M_{\odot}$ and an initial exo-Oort 
cloud distribution of semimajor 
axes of $a^{-3/2}$.  The simulation corresponds to the topmost (crimson) curve in the middle panel of Figure 
\ref{escape}.  The colours of the dots correspond to different values of $z$, with the highest values in red and 
the lowest in violet.  The blue patch of escaping comets at high $y$ and low $(x,z)$ occurs about 225 Myr along
the giant branch phase, well before any significant mass loss occurs.  This patch is due to a flyby star penetrating
to a closest-approach distance of just 394 au at that time.
}
\label{patchy}
\end{figure}

Each plot shows significant variation amongst some of the overlaid
histograms.  The most dramatic differences occur for the topmost
curves on the $a^{-1.5}$ and $a^{-2}$ plots, where nearly the entire
cloud escapes about 1 Gyr before the star becomes a WD, and when the
star still contains over 99.95\% of its original mass.  The sudden
escape in these systems is caused by a highly intrusive flyby at,
respectively, 225 Myr and 379 Myr after the start of the simulation.
The flybys penetrate to pericentre distances of 394 au and 528 au.
Figure \ref{patchy} is a correctly-scaled Cartesian plot of the points
of escape for all comets in the corresponding $a^{-1.5}$ simulation.
The dense blue patch in the upper-right portion of the ellipsoid is
the actual location where most comets escape during the intrusion.
This patch results in a highly inhomogeneous escape distribution.

We should actually expect such intrusions to occur with pericentres of
a few hundred au in a fraction of all systems.  Both \cite{zaktre2004}
and \cite{vermoe2012} demonstrate that one such close star-star
pericentre passage should occur on average once during a $\sim 10$ Gyr
main sequence lifetime.  Hence, we should expect a similarly
disruptive event during 10 Gyr of WD evolution.

\begin{figure}
\centerline{
\psfig{figure=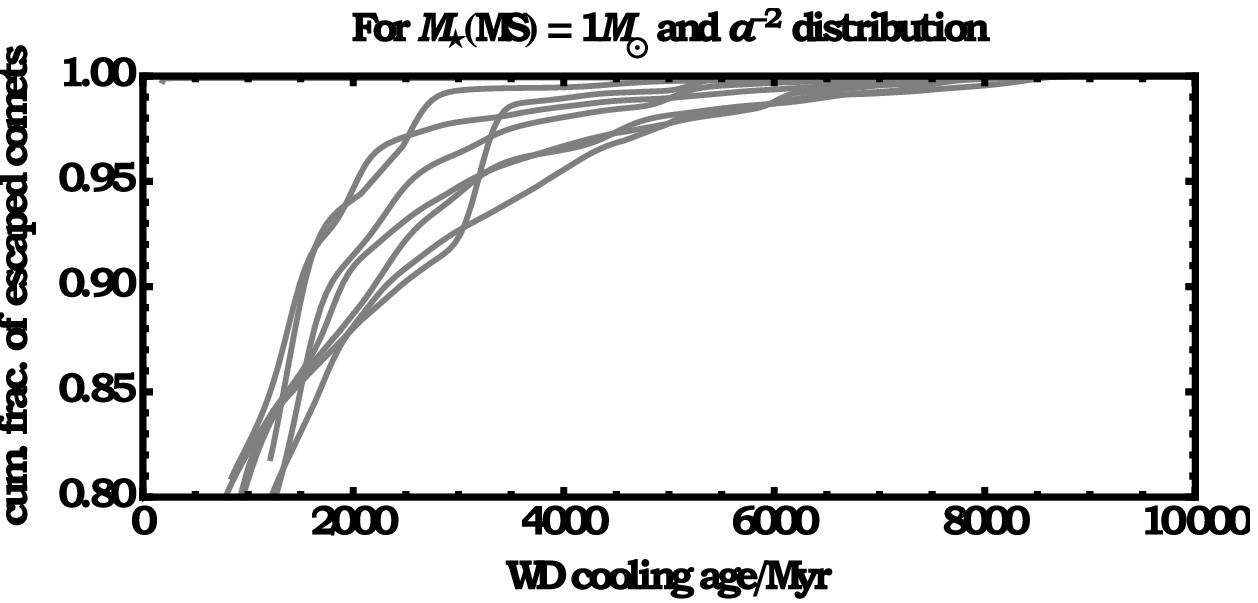,height=3.3cm} 
}
\centerline{ }
\centerline{
\psfig{figure=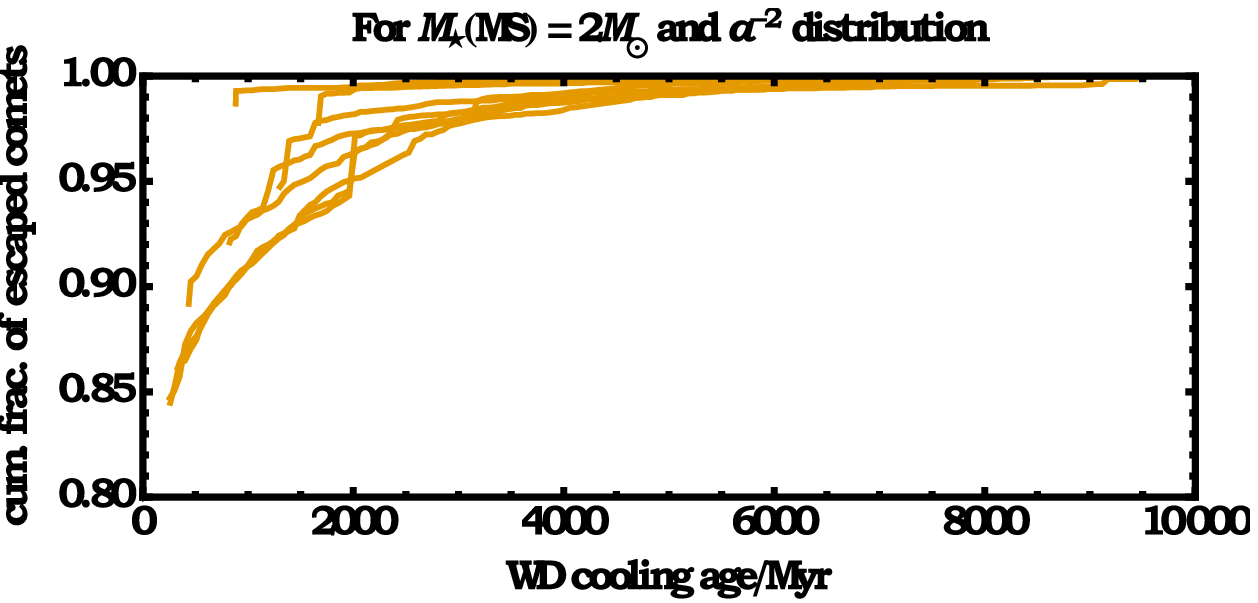,height=3.3cm} 
}
\centerline{ }
\centerline{
\psfig{figure=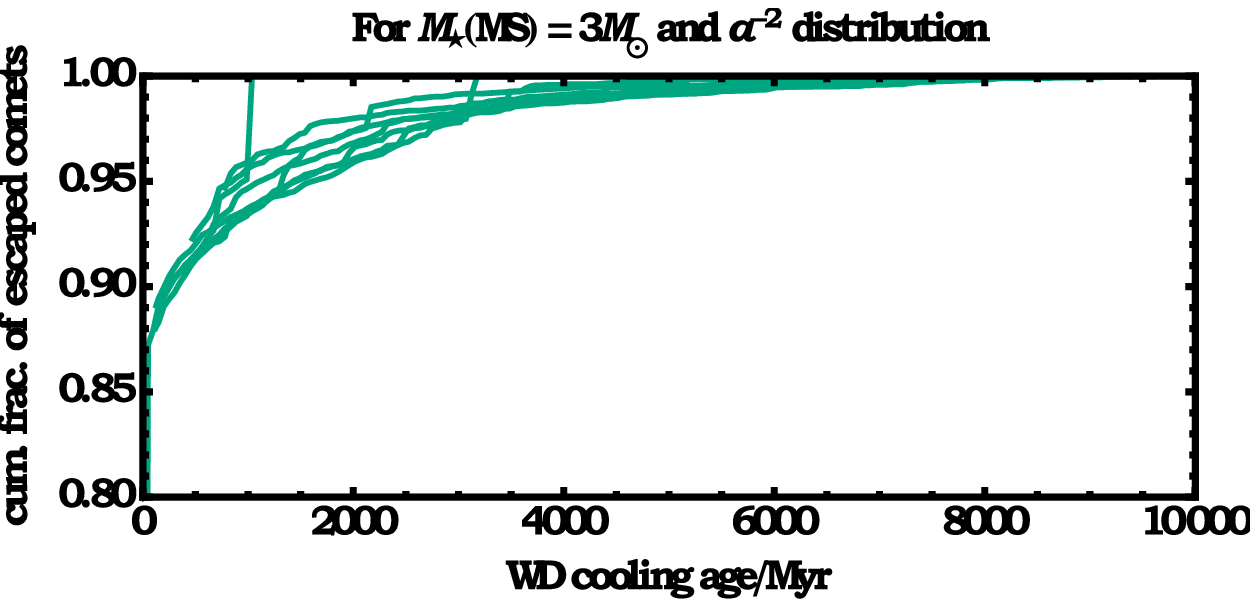,height=3.3cm} 
}
\centerline{ }
\centerline{
\psfig{figure=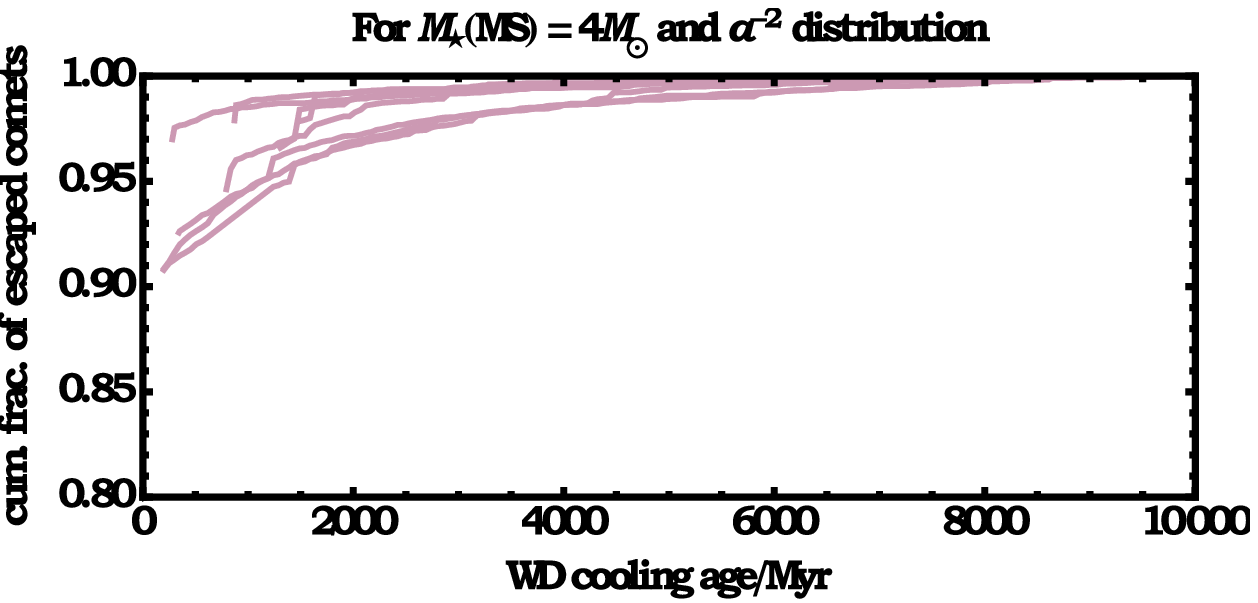,height=3.3cm} 
}
\centerline{ }
\centerline{
\psfig{figure=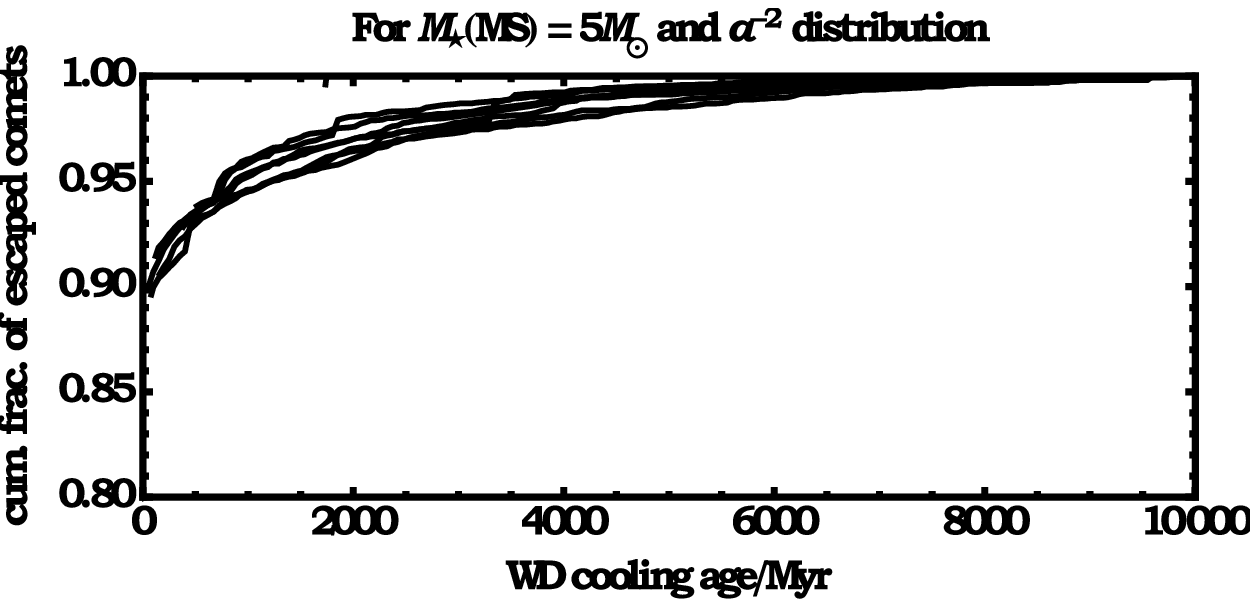,height=3.3cm} 
}
\caption{
Like Fig. \ref{escape}, except as a function of stellar mass.
Note that all y-axes here range from 80\%-100\%.
The figure demonstrates a steep dependence on stellar mass; for progenitor
masses greater than $3M_{\odot}$, all exo-Oort clouds are already at least
90\% depleted by the time the star becomes a WD.
}
\label{escape2}
\end{figure}

Although the lowest mass case is illustrative of the relevant
dynamics, the majority of currently observed WDs did not evolve from
Solar-mass main-sequence stars (see Fig. 1 of \citealt*{koeetal2014}).
Hence, determining how the reservoir of available exo-Oort cloud comets
is depleted as a function of stellar mass is important.  Figure \ref{escape2}
showcases this dependence for the representative $a^{-2}$ initial power-law 
distribution of exo-Oort cloud comets.  The dependence is steep, particularly
as the progenitor stellar mass is increased from $1M_{\odot}$ to $3M_{\odot}$.
Consequently, we expect the accumulation of H due to exo-Oort cloud comets
to be a function of both the WD mass itself, and its cooling age.

\section{Discussion}

\subsection{Observing cometary accretion}

In contrast to accretion of rocky material, an ongoing process which is currently
observed, we have not yet detected the accretion of cometary, volatile material 
onto WDs. Two reasons are (1)
entry into the WD Roche radius is potentially less frequent for 
exo-Oort cloud comets
than for asteroids, and (2) the different lifetimes of the gaseous
disc -- produced from the sublimation of a frozen iceball (comet) --
and the dusty disc, produced from the tidal disruption of a dry rocky
asteroid. The key parameter for the subsequent of the evolution of
these discs is their viscosity, which is much higher for the gaseous
discs compared than for dusty discs.  Dusty discs have an estimated
lifetime of $\sim 10^{4.5} - 10^{6.5}$ yr \citep{giretal2012} whereas
gaseous discs sustain themselves for only a few $10^3$ yr
\citep{metetal2012} (see also \citealt{jura2008} for estimates of the
gas/dust disc life times). Bearing in mind that about 30 WDs with
dust disc detections are currently known, it is unsurprising
that we have not yet identified an unambiguous case of
ongoing accretion from a tidally disrupted comet.

\subsection{Progenitors of metal pollutants}

Regarding sourcing metal pollution in WD atmospheres, the outcome of
this study emphasises the need to pursue other avenues of
investigation.  Considerations of the debris abundances aside, the
dynamics of injecting comets from exo-Oort clouds orbiting WDs into
the Roche radii of the WDs cannot explain the mass accumulated over
convective timescales in DBZ WDs.  The contribution of these comets to the
pollution is negligible unless we are underestimating the typical
masses of exo-Oort clouds by several orders of magnitude.

Other potential metal pollutant progenitors are planets, moons or asteroids.
Planets are unlikely because the frequency of direct collisions with the
WD is too low \citep{debsig2002,veretal2013a,voyetal2013,musetal2014}.
Asteroids and Kuiper-belt-like objects have also been the subject of recent
investigation.  \cite{bonetal2011} demonstrate that scattering between a
planet and an exo-Kuiper belt during the giant branch phases of evolution
can generate a sufficient quantity of mass in the inner planetary system to
match heavy metal accretion rates, but do not model how this mass obtains
WD-grazing orbits.  \cite{debetal2012} and \cite{freetal2014} show how
resonant diffusion between a planet and an exo-asteroid belt can represent an
effective mechanism to transfer asteroids to the WD Roche radius, but do
not perform simulations past 1 Gyr, and suggest that super-exo-asteroid
belt masses are required to explain the observations.
Further, a debris field generated by the rotational break-up of asteroids from
giant star radiation \citep{veretal2014c}
could affect the size distributions of potentially polluting bodies.
Sub-metre-sized particles could be dragged inward during giant branch envelope
evolution and be available as pollutants \citep{donetal2010}, but probably would not 
exist in sufficient quantities to account for the observed accreted mass.
Overall, none of theoretical models has yet to provide a complete 
explanation of the observations.

\subsection{Simulation subtleties}

Our results here are model-dependent, which is why we have taken a Monte-Carlo approach to 
our simulations.  The structure of an exo-Oort cloud at 
the start of the giant branch phases is unknown, 
as is the structure at the start of the main sequence.  Although all of our comets initially harboured
semimajor axes greater than $10^4$ au, mechanisms which produce Sedna and other inner Oort cloud 
objects rely on planets \citep[e.g.][]{gometal2006,glacha2006,lewetal2013}, or the 
Sun's birth environment \citep{braetal2006,levetal2010}, or migration within the galaxy 
\citep{kaietal2011}. These mechanisms can allow for speculation on whether the Sun's cloud is 
typical, but exploring them comprehensively is computationally prohibitive.

Our result is resolution-limited, as on average we see one WD Roche radius engulfment for 
every two simulations. However, including 5000 comets per simulation already pushes current computational 
resources.  We succeeded in integrating so many comets amidst mass loss for 10 Gyr because their wide orbits 
allow the integrator to use timesteps which are orders of magnitude larger than those from planetary 
simulations. Inserting planets in our simulations is not feasible because of the prohibitive timestep 
constraints they would introduce.  Further, they would force us to reduce the number of comets per simulation 
to the extent that we might not see any engulfment at all.

\subsection{The effect of planets}
As just mentioned, planetary influences may be a key factor in explaining accretion rates in systems with asteroid or Kuiper belts.  Their influence on exo-Oort clouds, however, is not as impactful.

We did not include planetary systems, as the computation demands of integrating short period orbits (and a larger number of massive bodies) is prohibitive.  However, Oort clouds are initially created by the gravitational scattering of planets \citep{oort1950}.  Such a planetary system might influence the evolution of the comet orbits near the star, altering the rate at which comets enter the Roche radius. \citet{horetal2010}, for instance, found that the flux of Oort cloud comets into the terrestrial planet region of the solar system is reduced by a factor of a few by Jupiter, compared to a system with a less massive planet in its place, as Jupiter can effectively eject incoming Oort cloud comets.  A similar result is found by \citet{lewetal2013}, who find that systems with a Saturn-mass planet or higher can reduce the rate Oort cloud comets get into the inner solar system by a factor of a few.  Saturn mass planets (and larger) occur around $\sim 10\%$~of Sun-like stars \citep{mayetal2011} in periods of less than $\sim 10$~years, while longer periods have not been effectively probed by any detection method, and $\sim 10\%$~of A-stars have planets of a few Jupiter masses at separations of 5-300 au \citep{vigetal2012}.  This potential depletion is mitigated by uncertainties about the survival of planetary systems.  Planetary systems may be unstable during the main-sequence lifetime of stars \citep{chaetal1996,jurtre2008}, a problem that can be compounded by the post-main-sequence evolution \citep{veretal2013a,musetal2014}, and can of course be lost to the star directly during post-main-sequence evolution \citep{musvil2012}.  The necessity that a planetary system be there to form an Oort cloud does not mean it must persist through the WD phase.  The overall uncertainty in our predictions from not including planetary systems is thus perhaps a factor of a few; in the worst case it may approach an order of magnitude.

\subsection{Other dynamical considerations}

Other effects not included in our simulations are stellar-comet tides nor general relativity.  For a single 
periastron passage, stellar tides would not affect whether the comet enters the Roche radius.  Further, these 
tides will not have time to act over multiple orbits because the orbit is likely to change, particularly at 
apastron, due to Galactic tides and stellar flybys.  General relativity will push the comet away from the 
star at closest approach on the order of $4$ km $\times (M_{\star}/M_{\odot})$ \citep{veras2014}, and so has 
a negligible effect in our simulations here.

The dynamical mechanisms discussed in this work are also applicable to exo-Oort clouds around
stars which undergo supernovae.  The only differences are that the rate of mass loss is much higher
(see discussion in Section 3.3.4 of \citealt*{veretal2011}), the total amount of mass loss may
be higher, and that the parent star will become 
either a neutron star or black hole.  \cite{hills1983} provides an extensive analysis of how 
the orbital elements of secondaries which orbit exploding stars change due to instantaneous 
mass loss.  In his Section III-d-iii, he describes the specific outcomes of comets, and estimates 
that only after several Gyr would passing stars drive down the pericentre of some surviving 
comets enough to hit the neutron star.

If in the future we are able to probe WDs outside of the Solar neighbourhood, then the
comet pollution hypothesis may be worth revisiting.  Most WDs in the Milky Way are between
the Solar neighbourhood and the Galactic bulge.  Galactic tides towards the bulge cause 
eccentricity oscillations with a greater frequency than those in the Solar
neighbourhood and are not dominated by vertical tides \citep{vereva2013}.  Further, towards the bulge, 
the stellar flybys are more frequent {\it and} the three Hill ellipsoid axes are all 
smaller than in the Solar neighbourhood for a given stellar mass.

\section{Summary}

Atmospheric hydrogen represents a permanent historical tracer of WD 
accretion activity. We have performed a dynamical analysis of 
post-main-sequence exo-Oort clouds to determine their contribution to 
the accreted circumstellar mass observed in
WD atmospheres.  We find that comets infrequently but gradually 
increase the H budget in WD atmospheres. Galactic tides, stellar flybys, 
and the location of the Solar
neighbourhood in the Milky Way all conspire to yield an exo-Oort cloud
engulfment fraction of $10^{-5}$.
One can use this value in conjunction with an assumed exo-Oort cloud mass and 
WD atmosphere heavy metal mass total (which can reach $10^{25}$ g) to 
speculate about the accretion history of these stars on a case-by-case basis.

\section*{Acknowledgements}

We thank the referee for a thorough and helpful review of the manuscript. 
We also thank Alan P. Jackson, Jay Farihi and Roberto Raddi for useful discussions.
The research leading to these results has received funding from the
European Research Council under the European Union's Seventh Framework
Programme (FP/2007-2013) / ERC Grant Agreement n. 320964 (WDTracer)
and n. 279973 (DEBRIS).  DV and BTG are funded under WDTracer, and AS
is funded under DEBRIS.

\label{lastpage}
\end{document}